\documentclass[pra,pacs,twocolumn]{revtex4}

\usepackage{amsfonts}
\usepackage{amsmath}
\usepackage{amssymb}
\usepackage{bm}
\usepackage[dvips]{color}
\usepackage{graphicx}
\usepackage{subfigure}

\newcommand{\hide}[1]{}

\begin{document}


\title{Three-Level Landau-Zener Dynamics}
\author{Y. B. Band$^{1}$ and Y. Avishai$^{2}$}
\affiliation{
$^{1}$Department of Chemistry, Department of Physics, Department of Electro-Optics, and the Ilse Katz Center for Nano-Science, \\
Ben-Gurion University, Beer-Sheva 84105, Israel\\
$^{2}$Department of Physics, and the Ilse Katz Center for Nano-Science, \\
Ben-Gurion University, Beer-Sheva 84105, Israel}


\begin{abstract}
We compute Landau-Zener probabilities for 3-level systems with a linear sweep of the uncoupled energy levels of the 3$\times$3 Hamiltonian matrix $H(t)$. Two symmetry classes of Hamiltonians are studied: For $H(t) \in$ su(2) (expressible as a linear combination of the three spin 1 matrices), an analytic solution to the dynamical problem is obtained in terms of the parabolic cylinder $D$ functions. For $H(t) \in$ su(3) (expressible as a linear combination of the eight Gell-Mann matrices), numerical solutions are calculated.  In the adiabatic regime, full population transfer is obtained asymptotically at large times, but at intermediate times, all three levels are populated, and St\"uckelberg oscillations can manifest from the occurrence of two avoided crossings. For the open system, (wherein interaction with a reservoir occurs), we numerically solve a Markovian quantum master equation for the density matrix with Lindblad operators that models interaction with isotropic white Gaussian noise. We find that St\"uckelberg oscillations are suppressed and that the decoherence cannot be modeled in terms of simple a exponential.
\end{abstract}

\pacs{05.40.Ca, 05.40.-a, 07.50.Hp, 74.40.De}

\maketitle

\section{Introduction} \label{sec:Intro}

The slow time evolution of quantum mechanical systems with discrete energy spectra is well described using adiabatic approximations wherein the energy eigenvalue crossings and/or pseudo-crossings (also referred to as avoided crossings) occur as the parameters used to describe the system are varied, e.g, due to the presence of time-dependent electric or magnetic fields.  A well known example of avoided level crossing is the 2-level Landau-Zener (LZ) problem \cite{Landau, Zener, Stueckelberg, Majorana}.  The LZ model was used in 1932 to theoretically model molecular pre-dissociation. Here we consider the 3-level LZ problem, e.g., a system having a time-dependent Hamiltonian whose matrix form is given by
\begin{equation}  \label{eq:ham}
H(t) = \hbar \begin{pmatrix} a \, t&\Delta&0\\ \Delta&0&\Omega \\ 0&\Omega&-a \, t \end{pmatrix}.
\end{equation}
The time-dependent Schr\"odinger equation, $i \hbar \dot{\bm \Psi} = H(t) {\bm \Psi}$, with ${\boldsymbol \Psi}(t)=[\psi_1(t),\psi_2(t),\psi_3(t)]^T$ 
specifies the unitary dynamics.
(In what follows we take $\hbar = 1$).
The 3-level LZ model is used to describe many physical systems.  For example, consider a three-level model for a system with a ladder level structure in which two transitions are driven by two lasers having constant amplitudes and detunings that vary linearly with time.  The driven transitions connect level 1 to level 2 and level 2 to level 3.  The 1-3 transition is not allowed by electric dipole selection rules.  The dressed state Hamiltonian matrix for such a system is given by Eq.~(\ref{eq:ham}), where the off-diagonal matrix elements are proportional to the classical external laser fields that drive the transitions.  As another example of LZ dynamics, consider the ground state of nitrogen vacancy centers in diamond, which is a spin 1 system \cite{Doherty_13, Ajisaka_16}, with the nitrogen vacancy placed in an external magnetic field directed along the $z$-axis of the nitrogen vacancy, and the strength of the field varies linearly with time.  In this case, the Hamiltonian is of the form given in Eq.~(\ref{eq:ham_su3}).

If $\Omega = \Delta$, the Hamiltonian in Eq.~(\ref{eq:ham}) belongs to su(2), i.e., the algebra of the group SU(2) \cite{Iachello}.  In its three-dimensional representation it can be written as $H(t)=a t S_z + \sqrt{2} \Delta S_x$, where $S_i$, $i = x, y, z$ are the 3$\times$3 spin-1 matrices.  In this case we obtain an analytic solution of the time-dependent Schr\"odinger equation and derive expressions for the probabilities $P_j(t) = \vert \psi_j(t) \vert^2$ for $j = 1,2,3$.  The analytic solution is also valid for any Hamiltonian $\tilde H(t)$ which is unitarily equivalent to $H(t)$.  For $\Omega \ne \Delta$, $H(t) \in$ su(3), and more generally, for linearly time-varying Hamiltonians $\tilde H(t) \in$ su(3) \cite{su2_su3}, e.g.,
\begin{equation}  \label{eq:ham_su3}
\tilde H(t)=(a t)/2 (\lambda_3 + \sqrt{3} \lambda_8) + \sum_{j=1,2,4,5,6,7} \Delta_j \lambda_j ,
\end{equation}
where $\lambda_j$ are the 3$\times$3 Gell-Mann matrices \cite{Gellmann}, analytic solutions are not known to us. Instead, we numerically solve the time-dependent density matrix equations $i \frac{\partial \rho}{\partial t} = [H(t), \rho(t)]$ and obtain the time-dependent probabilities which are given by the diagonal elements of the density matrix $\rho(t)$, $P_j(t) = \rho_{jj}(t)$ for $j = 1,2,3$ .  

We also consider the dynamics of the 3-level LZ `open system' (wherein the system is coupled to an environment) \cite{master_eq}.  To do so, we add a Lindblad term $- \Gamma \rho(t)$ to the density matrix equation,
\begin{equation}  \label{eq:dmeq}
i \, \partial \rho/\partial t = [H(t), \rho(t)] - \Gamma \rho(t) ,
\end{equation}
where $\Gamma$ is a Lindblad operator \cite{master_eq} ($\Gamma$ is a matrix when a matrix representation of this equation is used).  This formalism models Gaussian white noise affecting the 3-level system.  The 3-level LZ decoherence dynamics have a complicated temporal behavior arising from the multiple decay timescales (a maximum of 8 such timescales can be present for 3-level systems).

We note that work related to the 3-level LZ problem has been previously reported.  Examples of such work are Ref.~\cite{Carroll_Hioe}, where the authors derived an approximate formula for the long time behavior of the occupation probabilities, Ref.~\cite{Kiselev_13} which studies su(3) LZ interferometry, and Ref.~\cite{Parafilo_Kiselev} where the authors considered the 3-level LZ problem and Rabi oscillations in a periodically driven Cooper-Pair box (see also Ref.~\cite{Shevchenko_10} for a review of Landau-Zener-St\"uckelberg interferometry).  Moreover, multi-level LZ problems have also been studied \cite{Yurovsky_99, multilevel1, multilevel2, multilevel3, multilevel4, multilevel5, multilevel6,multilevel7, multilevel8, multilevel9}.  As noted in Ref.~\cite{Yurovsky_99}, one of the problems in trying to get analytic expressions for multi-level LZ problems is that counterintuitive transitions involving a pair of successive crossings can occur, in which the second crossing precedes the first one along the direction of motion.  This problem can arise for 3-level systems of the su(3) classification (see Sec.~\ref{sec:NR}).  One of the features that distinguish this work from the previous literature is that we have developed an analytic solution for the time-evolution of the 3-level su(2) classification.

The outline of this paper is as follows.  In Sec.~\ref{sec:AS} we present the analytic solution for the time-dependent Schr\"odinger equation for Hamiltonian (\ref{eq:ham}) with $\Omega = \Delta$. Section \ref{sec:NR} presents numerical results obtained for Hamiltonian (\ref{eq:ham}) for both cases $\Omega = \Delta$ and $\Omega \ne \Delta$. The open system dynamics is analyzed in Sec.~\ref{sec:OS} employing a master equation method with Lindblad operators.  Finally, Sec.~\ref{sec:S&C} contains a summary and conclusions.

\section{Analytic Solution for $\mbox{su(2)}$ Hamiltonians} \label{sec:AS}
 
Given the Hamiltonian $H(t)$ in Eq.~(\ref{eq:ham}) with $\Omega = \Delta$ [i.e., $H(t) \in$ su(2)], the time-dependent Schr\"odinger equation yields a set of three coupled equations for the components of ${\bm \Psi}(t)$:
\begin{eqnarray}   \label{eq:scheq}
&& i \dot{\psi}_1=a t \psi_1 + \Delta \psi_2 \label{eq:1a} \\
&& i \dot{\psi}_2=\Delta (\psi_1+ \psi_3) \label{eq:1b} \\
&& i \dot{\psi}_3=-a t \psi_3+ \Delta \psi_2. \label{eq:1c} 
\end{eqnarray}
We obtain an equation for $\psi_1(t)$ by eliminating $\psi_2$ and $\psi_3$.  First eliminate $\psi_2$ from (\ref{eq:1a}) and then substitute the result in (\ref{eq:1b}), etc.  After some algebra we find
\begin{equation} \label{eq:psi1}
\dddot{\psi}_1+\left(2 i a + (at)^2  +2  \Delta^2  \right)\dot{\psi}_1+ a^2 t \psi_1=0.
\end{equation}
The solution to Eq.~(\ref{eq:psi1}) is given in terms of the parabolic cylinder $D$ function \cite{Abramowitz}:
\begin{eqnarray}  \label{eq:psi1_sol}
\psi_1(t) &=& C_1 \left[D\left(-\frac{i \Delta^2}{2 a}, (-1)^{1/4} \sqrt{a} \, t\right)\right]^2  \\ \nonumber
  &+& C_2 D\left(-\frac{i \Delta^2}{2 a}, (-1)^{1/4} \sqrt{a} \, t\right)   \\ \nonumber
  &\times&  D\left(-1 + \frac{i \Delta^2}{2 a}, (-1)^{3/4} \sqrt{a} \, t\right)  \\ \nonumber
  &+& C_3  \left[D\left(-1 + \frac{i \Delta^2}{2 a}, (-1)^{3/4} \sqrt{a} \, t\right)\right]^2 .
\end{eqnarray}
The initial conditions, specified at large negative time $t_0$ for the three 
components of $\Psi(t)$ are:
\begin{equation} \label{eq:3a}
\psi_1(t_0)=1, \quad \psi_2(t_0)=\psi_3(t_0)=0 .
\end{equation}
From these constraints, initial conditions for $\psi_1(t)$ and its first and second derivatives are derived. Explicitly, from Eq.~(\ref{eq:1a}), we get
\begin{equation}
\dot{\psi}_1(t_0) = -i a t_0 ,
\end{equation}
and by differentiating Eq.~(\ref{eq:1a}) we find
\begin{equation} \label{eq:4a}
\ddot{\psi}_1(t_0) = -ia - [(a t_0)^2+\Delta^2].
\end{equation}
These initial conditions can be used to determine $C_1$, $C_2$ and $C_3$ in Eq.~(\ref{eq:psi1_sol}).  Thus, a closed form for the analytic solution with initial conditions (\ref{eq:3a}) is obtained.  However, it is too long to be displayed here.

With $\Omega \ne \Delta$, i.e., for $H(t) \in$ su(3), $\psi_1(t)$ satisfies the differential equation
\begin{eqnarray} \label{eq:psi1_gen}
\dddot{\psi}_1+\left[2 i a+ (at)^2  +  (\Delta^2+\Omega^2)  \right]\dot{\psi}_1 \\ \nonumber
+ \left[a^2 +ia(\Omega^2-\Delta^2 )\right] t \, \psi_1=0 .
\end{eqnarray}
Unfortunately, the solution to Eq.~(\ref{eq:psi1_gen}) is not known in terms of special functions.  Clearly, when $\Omega = \Delta$, Eq.~(\ref{eq:psi1_gen}) reduces to (\ref{eq:psi1}).

\section{Numerical Results for closed system dynamics}  \label{sec:NR}

\begin{figure}
\centering
\subfigure[]
{\includegraphics[width=0.7\linewidth]{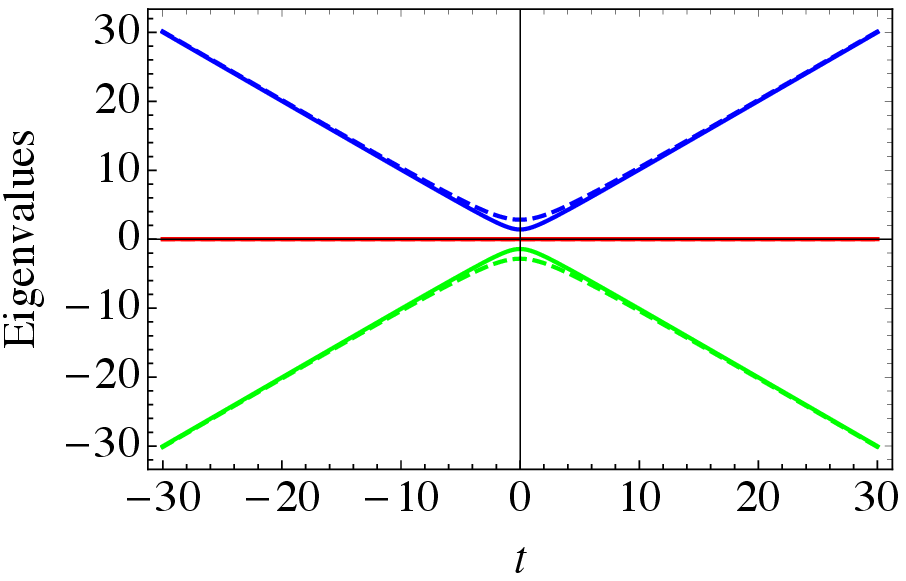}}
\centering
\subfigure[]
{\includegraphics[width=0.7\linewidth]{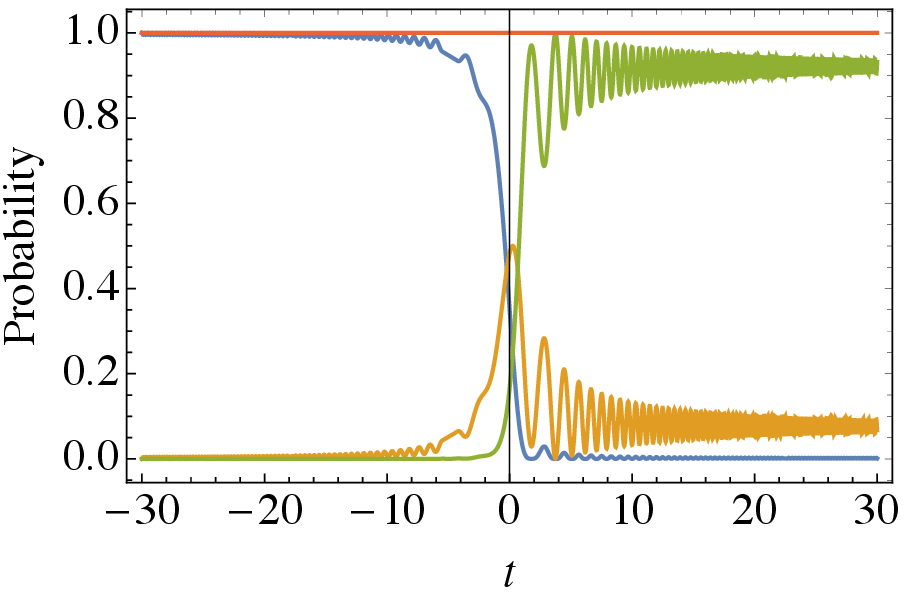}}
\centering
\subfigure[]
{\includegraphics[width=0.7\linewidth]{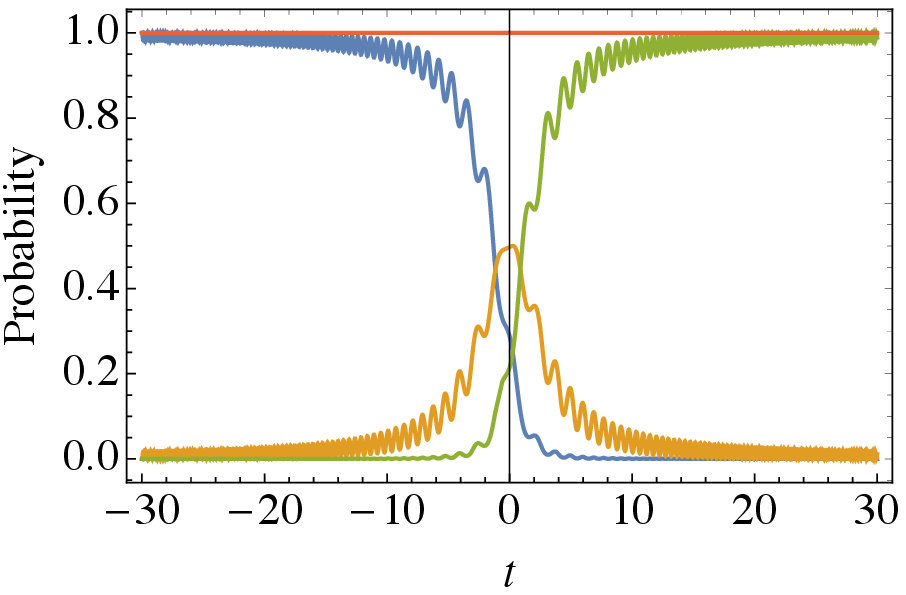}}
\caption{(a) Energy eigenvalues of the 3-level Hamiltonian versus time for $a = -1$, $\Omega = \Delta = 1$ shown as solid curves, and the eigenvalues for $\Omega = \Delta = 2$ shown as dashed curves [level 1 eigenvalue is the blue curve on top), 2 is the red curve in the middle and 3 is the green curve on bottom]. (b) Probability of levels 1 (blue curve on bottom right), 2 (orange curve in the middle right) and 3 (green curve on top right) versus time for $a = -1$, $\Omega = \Delta = 1$.  The blue curve $\rho_{11}(t)$ is initially unity at the initial time $t_0$, $\rho_{11}(t_0) = 1$.  (c) Probabilities $\rho_{11}$, $\rho_{22}$ and $\rho_{33}$ versus time for $a = 1$ and $\Omega = \Delta = 2$.  This case is nearly fully adiabatic, with population staying on the eigenvector with the highest eigenvalue, hence $\rho_{22} \to 1$ at large time.  The straight red line on top in (b) and (c) is the sum of the probabilities, ${\mathrm{Tr}}[\rho]$, which equals unity throughout the dynamics, as does the purity, ${\mathrm{Tr}}[\rho^2]$.}
\label{Fig_1}
\end{figure}

In this section we present results for the dynamics with no dissipation or decoherence.  These include both the su(2) ($\Omega=\Delta$) and the su(3) ($\Omega \ne \Delta$) dynamics.  Figure \ref{Fig_1} shows the eigenvalues of the Hamiltonian (\ref{eq:ham}) and the probabilities for levels 1, 2 and 3 versus time as obtained using the parameters $a = -1$ and $\Omega = \Delta$ (the units of $a$ are s$^{-2}$ and $\Delta$ are s$^{-1}$).  The energy eigenvalues are shown for $\Omega = \Delta = 1$ as solid curves and for $\Omega = \Delta = 2$ as dashed curves in Fig.~\ref{Fig_1}(a).  Clearly, as the off-diagonal coupling increases, the curves move farther apart, but, because of the symmetry in the coupling, the middle eigenvalue remains at zero energy.  The probabilities $P_j(t) = \rho_{jj}(t)$ for levels 1, 2 and 3 are plotted as a function of time for $\Omega = \Delta = 1$ in Fig.~\ref{Fig_1}(b), and for $\Omega = \Delta = 2$ in Fig.~\ref{Fig_1}(c).  At finite times, the probabilities undergo oscillations due to interference of fluxes arriving in a particular level at various times and/or to occurrence of more than one other level as shown in Fig.~\ref{Fig_1}(b) and (c).  The amplitude of oscillations shrinks with increasing coupling strength.  The population of level 2 builds up at intermediate times, but in the adiabatic limit [where $\Delta$ is large, see Fig.~\ref{Fig_1}(c)], the population of level 2 tends to zero at large time.  The numerical results using the density matrix equation (\ref{eq:dmeq}) with $\Gamma = 0$ fully agree with the results obtained using our analytic solution with the same initial conditions.

\begin{figure}
\centering
\subfigure[]
{\includegraphics[width=0.9\linewidth]{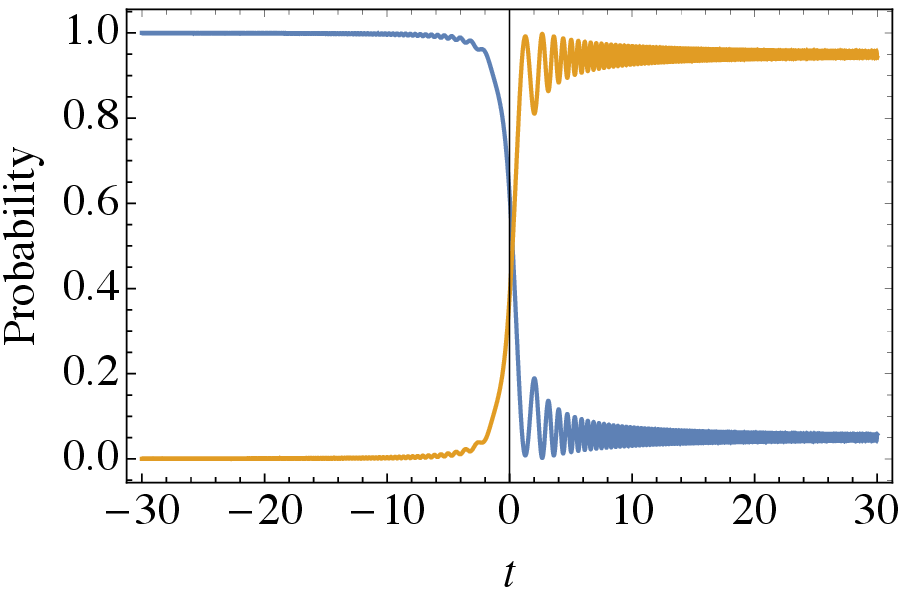}}
\centering
\subfigure[]
{\includegraphics[width=0.9\linewidth]{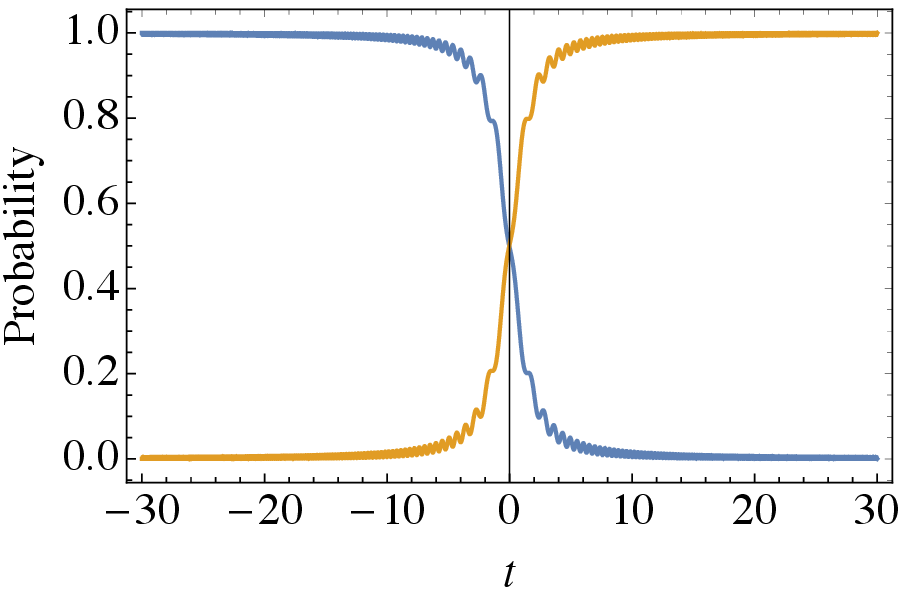}}
\centering
\subfigure[]
{\includegraphics[width=0.9\linewidth]{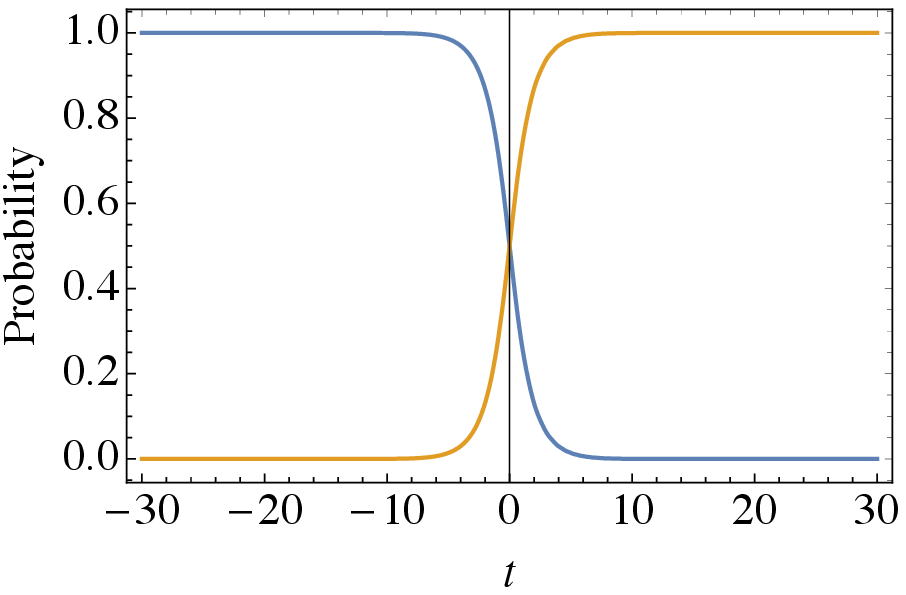}}
\caption{(a) Probability of levels 1 and 2 in the 2-level system versus time for $a = -1$ and $\Delta = 1$.  The initial state at early times is level 1 (blue curve).  (b) Probability of the levels versus time for $a = -1$ and $\Delta = 2$. (c) Probability with $a = -1$ and $\Delta = 2 \, e^{-(t/2\sigma)^2}$ with $\sigma = 5$.}
\label{Fig_2}
\end{figure}

For comparison, we recall the dynamics of the 2-level system governed by the 2$\times$2 Hamiltonian $H_{\rm{TLS}}(t) = \begin{pmatrix} a \, t&\Delta\\ \Delta &-a \, t\end{pmatrix}$ (see Ref.~\cite{Avishai_Band_14} for further details). The energy eigenvalues are the same as the non-zero energy eigenvalues of the 3-level problem with $\Delta=\Omega$ (multiplied by $1/\sqrt{2}$), see Fig.~\ref{Fig_1}(a).  Figure~\ref{Fig_2}(a) shows the probabilities $\rho_{11}(t)$ and $\rho_{22}(t)$ versus time for $\Delta = 1$ and Fig.~\ref{Fig_2}(b) for $\Delta = 2$, for which the evolution is approximately adiabatic.  The oscillations are completely suppressed by letting the off-diagonal coupling turn on and off as a Gaussian function of time, $\Delta(t) = 2 \, e^{-(t/2\sigma)^2}$, as shown in Fig.~\ref{Fig_2}(c), presumably because interference effects are thereby destroyed.

\begin{figure}
\centering
\subfigure[]
{\includegraphics[width=0.9\linewidth]{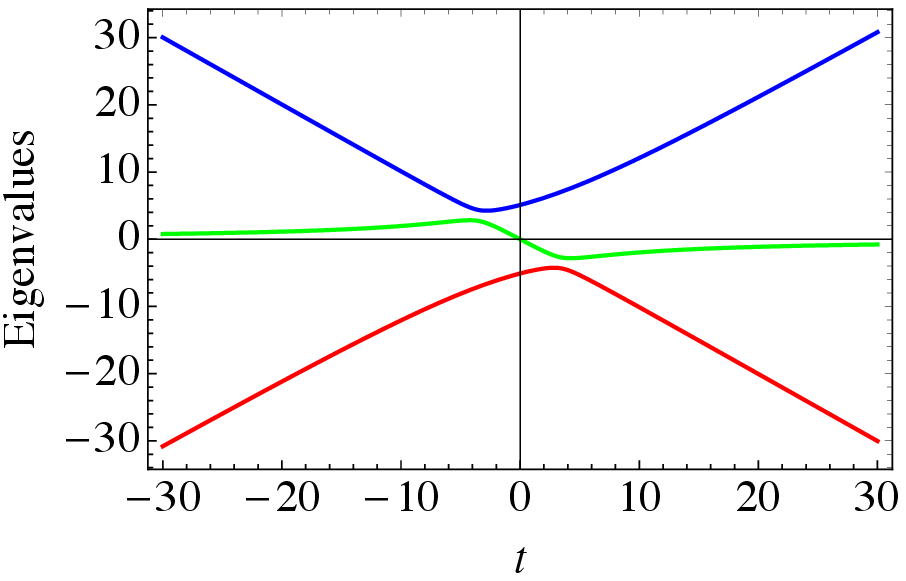}}
\centering
\subfigure[]
{\includegraphics[width=0.9\linewidth]{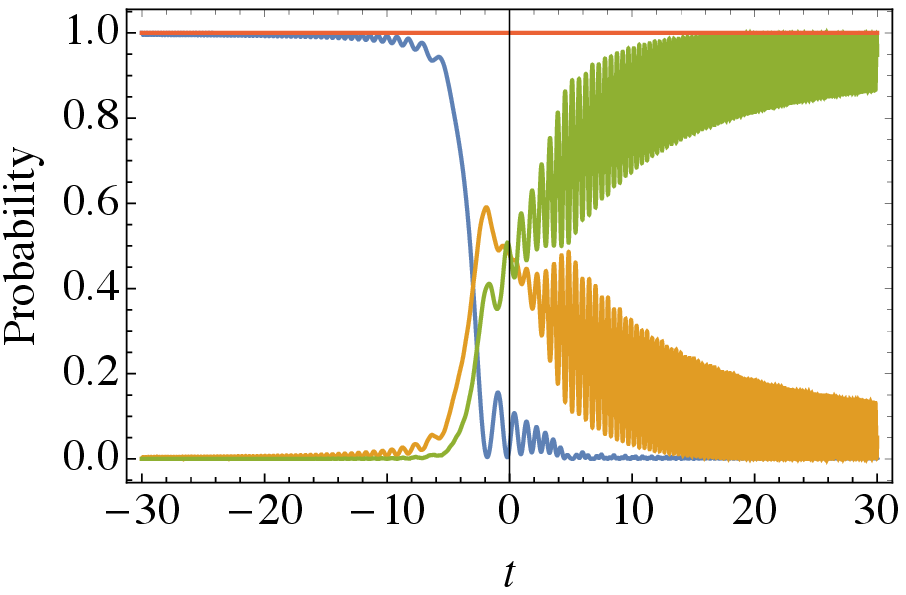}}
\caption{(a) Eigenvalues of the 3-level Hamiltonian versus time for $a = -1$, $\Delta = 1$ and $\Omega = 5$.  The level 1 eigenvalue is the blue curve on top, 2 is the red curve in the middle and 3 is the green curve on bottom.  (b) Probability of the levels 1, 2 and 3 versus time.  The straight red line on top is the sum of the probabilities, ${\mathrm{Tr}}[\rho]$, which remains unity during the dynamics, as does the purity ${\mathrm{Tr}}[\rho^2]$.}
\label{Fig_3}
\end{figure}

The dynamics of the 3-level LZ problem with $\Omega \ne \Delta$ can show pseudo-crossings which, under certain conditions, are similar to that of the 2-level LZ system [see Fig.~\ref{Fig_3}(a), where pseudo-crossing occurs twice, near $t = -4$ and $t = 4$].  However, taking the product of Landau-Zener amplitudes yield inaccurate probabilities because counterintuitive transitions can also play a role \cite{Yurovsky_99}. Figure~\ref{Fig_3} shows the eigenvalues and probabilities versus time for $a = -1$, $\Delta = 1$ and $\Omega = 5$. The energy eigenvalues are plotted in Fig.~\ref{Fig_3}(a), and the probabilities $\rho_{11}(t)$, $\rho_{22}(t)$ and $\rho_{33}(t)$  are shown in Fig.~\ref{Fig_3}(b).  For the 3-level system, St\"uckelberg oscillations can occur due to interference of amplitude flux arriving in a particular level via the avoided crossings at different times even for a linear energy sweep (in 2-level LZ dynamics, multiple avoided crossings occur only with non-linear sweeps).  Note that at large times, a large fraction of the probability initially in level 1 is transferred to level 3.

\section{Open System Dynamics} \label{sec:OS}

An open system is one that interacts with its environment (also referred to as a bath). Open systems undergo dephasing and decoherence.  There are several methods for modelling dynamics of open systems, including master equations \cite{master_eq, vanKampenBook}, the Monte Carlo wave-function approach \cite{Molmer_93}, and stochastic differential equation techniques \cite{vanKampenBook}.  Here we model dephasing and decoherence using a von-Neumann Liouville equation for the density matrix of the system with Lindblad operators \cite{Avishai_Band_14, master_eq, vanKampenBook}.  
For systems that are coupled to Gaussian white noise,  the stochastic dynamics can be described using the Schr\"{o}dinger--Langevin equation \cite{vanKampenBook}.  For a single white noise source one obtains the equation
\begin{equation}  \label{Schr_Langevin}
    i {\dot \psi} = H(t) \psi + \xi_0 \xi(t) {\cal V} \psi -
    \frac{\xi_0^2}{2} {\cal V}^{\dag} {\cal V} \psi ,
\end{equation}
where ${\cal V}$ are Lindblad operators, $\xi(t) = d w(t)/dt$, where $w(t)$ is a Wiener process, and the term proportional to $\xi_0^2$ insures unitarity \cite{vanKampenBook}.  Equation (\ref{Schr_Langevin}) can be generalized to include sets of Lindblad operators ${\cal V}_j$, sets of stochastic processes $w_j(t)$ and volatilities $\xi_{0,j}$, to obtain the general Schr\"{o}dinger--Langevin equation,
\begin{equation}  \label{gen_Schr_Langevin}
    {\dot \psi} = -i H(t) \psi + \sum_j \left(\xi_{0,j} \xi_j(t) {\cal V}_j
    - \frac{\xi_{0,j}^2}{2} {\cal V}^{\dag}_j {\cal V}_j \right) \psi.
\end{equation}
One could solve the stochastic equations (\ref{gen_Schr_Langevin}) to obtain the average and standard deviation of the probabilities reported below for the 3-level LZ problem, but this will take us a bit too far afield. Instead, we concentrate on the average over the stochasticity, which can be obtained from the Markovian quantum master equation for the density matrix $\rho(t)$ with Lindblad operators ${\cal V}_j$ \cite{master_eq, vanKampenBook}:
\begin{eqnarray}  \label{Eq:master}
    {\dot \rho} &=& -i [H(t), \rho(t)] \\ \nonumber 
    &+& \frac{1}{2} \sum_j \xi_{0,j}^2
    \left(2{\cal V}_j \rho(t) {\cal V}^{\dag}_j - \rho(t) {\cal
    V}^{\dag}_j {\cal V}_j - {\cal V}^{\dag}_j {\cal V}_j \rho(t)
    \right) .
\end{eqnarray}
In our case, we take the Lindblad operators to be the three spin-1 operators, ${\cal V}_j = S_j$, $j = x, y, z$, to model isotropic white noise.  

\begin{figure} [ht]
\centering
\includegraphics[width=0.9\linewidth]{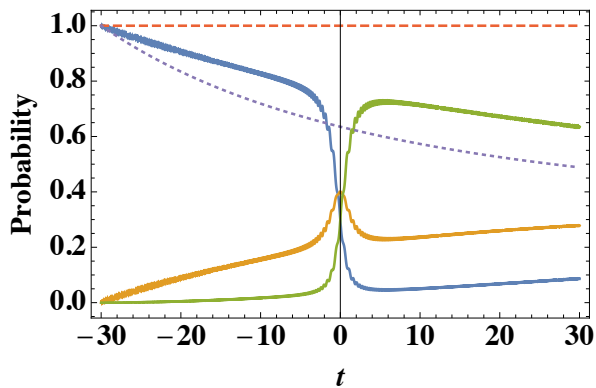}
\caption{Probabilities $\rho_{11}$ (blue curve on top left), $\rho_{22}$ (orange curve) and $\rho_{33}$ (green curve on top right) versus time for $a = 1$ and $\Omega = \Delta = 2$ [same parameters as in Fig.~\ref{Fig_1}(c)] with isotropic decay ($p_0 = 0.1$).  The red dashed line shows the sum of the probabilities, ${\mathrm{Tr}}[\rho]$, which remains equal to unity throughout the dynamics, and the purple dotted curve shows the purity, ${\mathrm{Tr}}[\rho^2]$, which uniformly decreases from unity as a function of time and goes asymptotically to 1/3 at large times.  Compare with Fig.~\ref{Fig_1}(c) for the unitary (no decay) case.}
\label{Fig_4}
\end{figure}

Figure \ref{Fig_4} shows the occupation probabilities $\rho_{11}$, $\rho_{22}$ and $\rho_{33}$ versus time when the volatilities are chosen such that $\xi_{0,j} = 0.1$ for $j = 1, 2, 3$.  The decoherence is apparent in each of the probabilities. The amplitudes of the St\"uckelberg oscillations decay with time as well.  At long time, the population is distributed among all three levels, as shown in Fig.~\ref{Fig_4}.  The total probability $\sum_{j=1}^3 \rho_{jj}(t)$ remains equal to 1 (see red dashed line in Fig.~\ref{Fig_4}) but the purity ${\mathrm{Tr}}[\rho^2]$ decays to 1/3 (see purple dotted curve), and the decoherence is {\em not} as simple as a single exponential decay, $e^{-(t - t_0)/\tau}$. Rather, a more complicated temporal dependence ensues.   The complicated decay can be understood as follows. For a time-independent Hamiltonian, each of the matrix elements of the density matrix can be expressed as
$$
\rho_{\alpha\beta}(t) =a^{\alpha,\beta}_0 +\sum_{i=1}^8 a_i^{\alpha,\beta} \exp (\lambda_i t) ,\ \ 
(\alpha,\beta=1,2,3)
$$
where the $\lambda_i$ ($i = 0, 1, \ldots, 8$) are the 9 eigenvalues of the 9$\times$9 Liouvillian operator, the real parts of $\lambda_i$ determine decay rates and the imaginary parts determine energy eigenvalue differences, $a_i$ ($i = 1, \ldots, 8$) are the amplitude coefficients, the coefficient $a_0$ corresponds to the amplitude of the steady state whose existence is guaranteed by trace preservation, and $\lambda_0$ is zero.  Hence, for a 3-level system, the maximum number of possible timescales that determine the population decay and the coherence dynamics is 8 (the number of non-zero eigenvalues), but there may be a lower the number due to symmetry. For a time-dependent Hamiltonian, in the adiabatic regime, the eigenvalues $\lambda_i$ and amplitudes $a_i$ are time-dependent, and an adiabatic expansion can still be carried through \cite{Band_92}.  In any case, one decay rate of the populations and the purity is in general not enough for a 3-level system.

\section{Summary and Conclusions} \label{sec:S&C}

We developed an analytic solution to the 3-level LZ problem for the Hamiltonian in Eq.~(\ref{eq:ham}) with $\Omega = \Delta$.  The solution [see Eq.~(\ref{eq:psi1_sol})] is given in terms of the parabolic cylinder $D$ functions \cite{Abramowitz}. This analytic solution holds for any Hamiltonian $H(t)$ with a linear sweep that can be written as a linear combination of the 3$\times$3 spin-1 matrices, $S_x, S_y, S_z$, i.e., $H(t) \in$ su(2).  We also calculate the dynamics for the case $\Omega \ne \Delta$, wherein the eigenvalues of the Hamiltonian may  have two separate pseudo-crossings.  This Hamiltonian belongs to su(3) and can be expanded as a linear combination of the Gell-Mann matrices \cite{su2_su3, Gellmann}.  When the sweep-rate $a$ is small and the coupling(s) $\Delta$ (and $\Omega$) is (are) large, the evolution is adiabatic and the system stays on the initial eigenvector, but even in the adiabatic limit, interference oscillations are present at intermediate times.  For the su(3) case, the physics of the LZ transitions involves two time-separated pseudo-crossings (two avoided crossings occurring at different times), as shown in Fig.~\ref{Fig_3}(a), but the calculation of the transition probabilities needs to be carried out as a 3$\times$3 matrix problem \cite{Yurovsky_99}.  We also numerically solved the open system problem using the Markovian quantum master equation for the density matrix $\rho(t)$ with Lindblad operators ${\cal V}_j = S_j$ to model isotropic Gaussian white noise \cite{Avishai_Band_14}.  In the presence of such noise, St\"uckelberg oscillations are suppressed due to the decay associated with fluctuations (recall the fluctuation dissipation theorem \cite{Band_Avishai}).  Moreover, the decoherence cannot be described by a single exponential; it is characterized by a more complicated function of time due to the presence of the multiple decay timescales of 3-level systems (a maximum of 8 such timescales occur).  We note that open system LZ dynamics may involve other than Gaussian white noise, including colored Gaussian noise \cite{Kenmoe_13} or other types of noise that lead to non-Markovian dynamics, but we have not addressed these issues here.

\begin{acknowledgments}
This work was supported in part by grants from the DFG through the DIP program (FO703/2-1).
\end{acknowledgments}



\begin{thebibliography}{99}
\bibitem{su2_su3}
3-level Hamiltonians can be classified according to whether they can be expanded in terms of su(2) generators (the 3$\times$3 spin-1 matrices $S_x$, $S_x$ and $S_z$) or in terms of the more general su(3) generators given by the 3$\times$3 Gellman matrices $\lambda_j$, $j = 1, \ldots 8$.  The relevance of the group structure in LZ physics was first noted in Refs.~\cite{Kiselev_13, Parafilo_Kiselev}.
\end{thebibliography}
\end{document}